\documentclass[aps,12pt,nofootinbib,showpacs,showkeys,preprintnumbers,amsmath,amssymb]{revtex4-1}

\usepackage{amsmath,amsfonts,amssymb,amsthm,bm,bbm,bbold,braket,color,dsfont,fancybox,hyperref,
srcltx,subfigure,ucs,yfonts,cancel,xfrac,verbatim,xcolor}
        
\usepackage{float}

\usepackage[section]{placeins}

\usepackage{mathtools}
\newcommand{\beq}{\begin{equation}}
\newcommand{\eeq}{\end{equation}}
\newcommand{\bea}{\begin{eqnarray}}
\newcommand{\eea}{\end{eqnarray}}
\newcommand{\beas}{\begin{eqnarray*}}
\newcommand{\eeas}{\end{eqnarray*}}
\newcommand{\bi}{\begin{itemize}}
\newcommand{\ei}{\end{itemize}}

\usepackage{amsmath,amsfonts,amssymb,amsthm,bm,bbm,bbold,braket,color,dsfont,fancybox,hyperref,
srcltx,slashed,ucs,yfonts,cancel,xfrac,verbatim,xcolor}

\usepackage[inline,shortlabels]{enumitem}
\usepackage{nicefrac}
\usepackage{multirow}

\DeclareMathAlphabet{\mathpzc}{OT1}{pzc}{m}{it}
\definecolor{gold}{rgb}{1,0.8,0}
\definecolor{nara}{rgb}{1,0.4,0.1}
\definecolor{goldo}{rgb}{1,0.7,0}
\definecolor{greeno}{rgb}{0,0.8,0}
\def\bes{\begin{subequations}}
\def\ees{\end{subequations}}
\def\be{\begin{equation}}
\def\ee{\end{equation}}
\def\bea{\begin{eqnarray}}
\def\eea{\end{eqnarray}}
\def\ba{\begin{eqnarray}}
\def\ea{\end{eqnarray}}
\def\bear{\begin{array}}
\def\eear{\end{array}}

\newcommand{\bpm}{\begin{pmatrix}}
\newcommand{\epm}{\end{pmatrix}}
\newcommand{\BM}{\left(\begin{array}}		
\newcommand{\BMC}{\left[\begin{array}}		

\newcommand{\EM}{\end{array}\right)}		
\newcommand{\EMC}{\end{array}\right]}		

	% ...

\newcommand{\com}[1]{}
\newcommand{\K}{\mathcal{K}}

\begin{document}
\begin{flushright}
\end{flushright}

\title{Discovering heavy neutrino oscillations in rare $B_c^{\pm}$ meson decays at HL-LHCb} 

\author{Sebastian Tapia$^{1}$}
\email{s.tapia@cern.ch}
\author{Marcelo Vidal-Bravo$^{2,3}$}
\email{M.vidalbravo@uandresbello.edu}
\author{Jilberto Zamora-Sa\'a$^{2,3}$}
\email{jilberto.zamora@unab.cl} \email{jilberto.zamorasaa@cern.ch}

\affiliation{$^1$Department of Physics and Astronomy, Iowa State University, USA.}
\affiliation{$^2$Center for Theoretical and Experimental Particle Physics, Facultad de Ciencias Exactas, Universidad Andres Bello, Fernandez Concha 700, Santiago, Chile}
\affiliation{$^3$Millennium Institute for Subatomic physics at high
energy frontier - SAPHIR, Fernandez Concha 700, Santiago, Chile.}

\begin{abstract}
In this work, we study the lepton flavor and lepton number violating $B_{c}$ meson decays via two intermediate on-shell Majorana neutrinos $N_j$ into two charged leptons and a charged pion $B_{c}^{\pm} \to \mu^{\pm}  \ N_j \to \mu^{\pm} \tau^{\pm} \pi^{\mp}$. We evaluated the possibility to measure the modulation of the decay width along the detector length produced as a consequence of the lepton flavor violating process, in a scenario where the heavy neutrinos masses range between $2.0$ GeV $\leq M_N \leq 6.0$ GeV. We study some realistic conditions which could lead to the observation of this phenomenon at futures $B$ factories such HL-LHCb.

\end{abstract}

\maketitle

\section{Introduction}
\label{s1}
The first indications of physics beyond the standard model (SM) come from the baryonic asymmetry of the universe (BAU), dark matter (DM), and neutrino oscillations (NOs).  In last decades NOs experiments have shown that active neutrinos ($\nu$) are very light massive particles $M_{\nu} \sim 1$eV~\cite{Fukuda:1998mi,Eguchi:2002dm} and, consequently, the SM is not a final theory, and must be extended. There are several SM extensions, that allow explaining the small active neutrino masses, however, in this paper we pay attention to those based on the See-Saw Mechanism (SSM) \cite{Mohapatra:2005wg,Mohapatra:2006gs}. The SSM introduces a new Heavy Majorana particle (singlet under $SU(2)_L$ symmetry group), commonly called Heavy Neutrino (HN), which by means of inducing a dim-5 operator \cite{Weinberg:1979sa} leads to a very light active Majorana neutrino. These newly introduced HNs have a highly suppressed interaction with gauge bosons ($Z,W^{\pm}$) and leptons ($e, \mu, \tau$), making its detection a challenging task. However, although this suppression, the existence of HNs can be explored via rare meson decays~\cite{Zhang:2020hwj,Abada:2019bac,Drewes:2019byd,Godbole:2020jqw,Dib:2000wm,Cvetic:2012hd,Cvetic:2013eza,Cvetic:2014nla,Cvetic:2015naa,Cvetic:2015ura,Dib:2014pga,Zamora-Saa:2016qlk,Milanes:2018aku,Mejia-Guisao:2017gqp,Cvetic:2020lyh}, colliders~\cite{Das:2018usr,Das:2017nvm,Das:2012ze,Antusch:2017ebe,Das:2017rsu, Das:2017zjc,Chakraborty:2018khw,Cvetic:2019shl,Antusch:2016ejd,Cottin:2018nms,Duarte:2018kiv,Drewes:2019fou,Dev:2019rxh, Cvetic:2018elt,Cvetic:2019rms,Das:2018hph,Das:2016hof,Das:2017hmg,Milanes:2016rzr}, and tau factories \cite{Zamora-Saa:2016ito,Tapia:2019coy,Kim:2017pra,Dib:2019tuj}.

One of the most promising SM extensions based on SSM is the Neutrino-Minimal-Standard-Model ($\nu$MSM)~\cite{Asaka:2005an,Asaka:2005pn}, which introduces two almost degenerate HN's with masses $M_{N1} \approx M_{N2} \sim 1 $GeV, and a third HN with mass $M_{N3} \sim $keV which is a natural candidate for DM. Apart to explain the small active neutrino masses, the $\nu$MSM allows to explain sucesfully the BAU by means of \emph{leptogenesis from HNs oscillations}, also known as Akhmedov-Rubakov-Smirnov (ARS) mechanism~\cite{Akhmedov:1998qx}.

In a previous article~\cite{Cvetic:2015ura}, we have described the effects of Heavy Neutrino Oscillations (HNOs) in the so-called rare Lepton Number Violating (LNV) and Lepton Flavor Violating (LFV) pseudoscalar $B$ meson decays, via two almost degenerate heavy on-shell Majorana neutrinos ($M_{N_i} \sim 1$GeV), which can oscillate among themselves.  The aim of this article is to develop a more realistic analysis of the experimental conditions needs to detect the aforementioned phenomenon. We will focus specially on the HL-LHCb which due to his excellent detector resolution \cite{LHCb:2021moh,Aaij:2020buf} could make possible the observation of the HNOs.  Similar studies have been performed for other experiments(see Refs. \cite{Cvetic:2018elt,Tapia:2019coy, Cvetic:2020lyh,Tastet:2019nqj}).

The work is arranged as follows: In Sec.~\ref{s3}, we study the production of the heavy neutrinos in $B_c^{\pm}$ meson decays. In Sec.~\ref{sec:simu}, we describe the simulations of the HN production. In Sec.~\ref{sec:dis}, we present the results and a discussion of its and In sec.~\ref{summ} we provide a brief summary of the article.

\section{Production of the RHN}
\label{s3}
\begin{figure}[hbt]
\centering
\includegraphics[scale = 0.7]{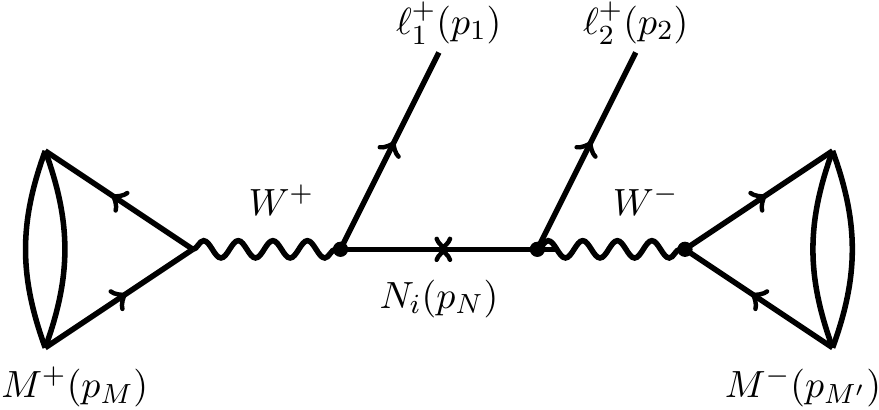}\hspace{0.3 cm}
\includegraphics[scale = 0.7]{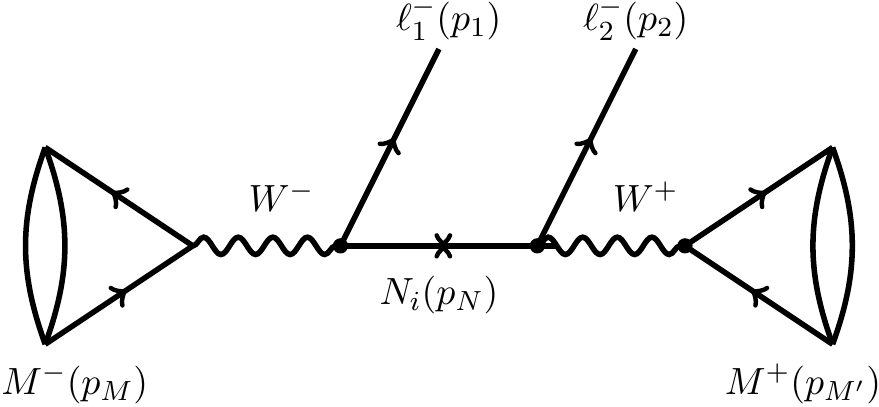}
\caption{The $M^{\pm}$ pseudoscalar meson decays, intermediated by Heavy Neutrinos. Left Panel: Feynman diagrams for the LFV and LNV process $M^+\rightarrow \ell_1^+ \ell_2^+ \pi^{-}$. Right Panel: Feynman diagrams for the LFV and LNV process $B^-\rightarrow \ell_1^- \ell_2^- \pi^{+}$. We remarks that in this study we will consider $M=B_c$, $M'=\pi$, $\ell_1=\mu$ and $\ell_2=\tau$.}
\label{fig:feynn}
\end{figure}

As we stated above, we are interested in studying the lepton flavor and leptop number violation processes ($B_{c}^{\pm} \to \mu^{\pm}  \ N_j \to \mu^{\pm} \tau^{\pm} \pi^{\mp}$) which are caracterized by the following Feynman diagrams (Fig.~\ref{fig:feynn}). 
In this work we will consider the scenario where the two heavy neutrino ($N_1$ and $N_2$) masses fall in the range of a few GeVs and are almost degenerate ($M_{N_1} \approx M_{N_2}$).

The mixing coefficient between the standard flavor neutrino $\nu_{\ell}$ ($\ell = e, \mu, \tau$) and the heavy mass eigenstate $N_i$ is   $B_{\ell N_i}$ (i = 1, 2), then the light neutrino flavor state can be defined as

\begin{equation}
\nu_{\ell} = \sum_{i=1}^{3} B_{\ell \nu_i} \nu_i +\quad  \underbrace{(B_{\ell N_1} N_1 + B_{\ell N_2} N_2)}_{\rm Heavy\ Neutrino\ Sector}\ ,
\end{equation}
where $B_{\ell \nu_i}$ (i = 1, 2, 3) and $B_{\ell N_j}$ (j = 1, 2) are the complex elements of the $5\times5$ PMNS matrix, and will be parameterized as follow

\begin{equation}
B_{\ell \nu_i} = |B_{\ell \nu_i}| e^{i \theta_{\ell i}}\ , (i =1,2,3) \quad {\rm and} \quad B_{\ell N_j} = |B_{\ell N_j}| e^{i \theta_{\ell N_j}}\ ,(j=1,2)\ .
\end{equation}

\begin{figure}[hbt]
\centering
\includegraphics[scale = 0.65]{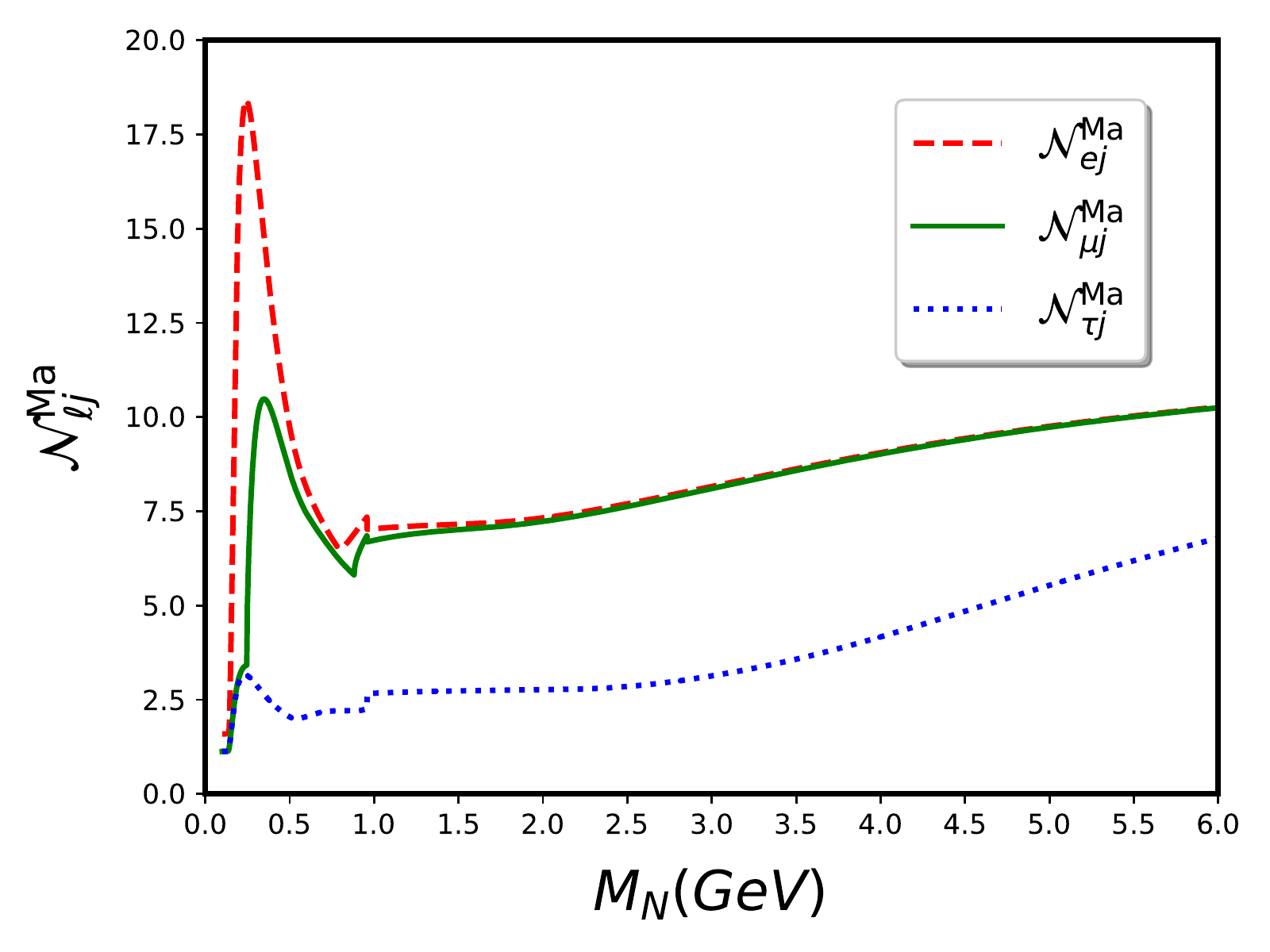}
\caption{Effective mixing coefficients ${\cal N}_{\ell j}^{\rm Ma}$ for Majorana neutrinos. Data taken from \cite{Atre:2009rg}.}
\label{fig:efcoef}
\end{figure}

 The mass difference between HNs is expressed as ($|\Delta M_N|=M_{N2}-M_{N1}\equiv Y \Gamma_N$), where $Y$ stand to measures the mass difference in terms of $\Gamma_{N} = (1/2)(\Gamma_{N_1} + \Gamma_{N_2})$ which is the (average of the) total decay width of the intermediate Heavy Neutrino. The decay width $\Gamma_{\rm Ma}(M_{N_i})$ of a single Heavy Neutrino $N_i$ is
\begin{equation}
  \Gamma_{\rm Ma}(M_{N_i}) \equiv \Gamma_{N_i}   \approx  \K_i^{\rm Ma}\ \frac{G_F^2 M_{N_i}^5}{96\pi^3} 
\label{DNwidth}
\end{equation}
 where
 \begin{equation} 
 \K_i^{\rm Ma} = {\cal N}_{e i}^{\rm Ma} \; |B_{e N_i}|^2 + {\cal N}_{\mu i}^{\rm Ma} \; |B_{\mu N_i}|^2 + {\cal N}_{\tau i}^{\rm Ma} \; |B_{\tau N_i}|^2,
\label{DNwidth1}
\end{equation}
here the factors $B_{\ell N}$ are the heavy-light mixing elements of the PMNS matrix\footnote{In this work we define the light neutrino flavor state as $\nu_{\ell}=\sum_{i=1}^{3}U_{\ell i} \nu_i + \sum_{j=1}^2$ $B_{\ell N} N_j$. However, other authors also use $U_{\ell N}$ or $V_{\ell N}$ as the heavy-light mixings elements  (i.e. $B_{\ell N} \equiv  U_{\ell N} \equiv  V_{\ell N}$).} and ${\cal N}_{\ell i}^{\rm Ma}$ are the effective mixing coefficients which account for all possible decay channels of $N_i$ and are presented in Fig.~\ref{fig:efcoef} for our $M_N$ range of interest ($0 \leq M_N \leq 6$ GeV).

It is important to mention that due to the dependency on $|B_{\ell N_i}|$ the factors $\mathcal {K}^{\rm Ma}_i$ could be in principle different for $N_1$ and $N_2$, it means, it is possible that $|B_{\ell N_1}| \neq |B_{\ell N_2} $, and consequently $\mathcal {K}^{\rm Ma}_1$ dominates over $ \mathcal {K}^{\rm Ma}_2 $ or vice versa. The factor $\mathcal {K}^{\rm Ma}_i $ only appear in $ \Gamma_N(M_{N_i})$ (Eq. \ref{DNwidth}), all our numerical calculations have been performed for  $\Gamma_{N} = (1/2)(\Gamma_{N_1} + \Gamma_{N_2})$, i.e. $ \Gamma_N(M_N) \approx \Gamma_N (M_{N_1}) /2 $ if $ \mathcal {K}^{\rm Ma}_1 \gg  \mathcal {K}^{\rm Ma}_2 $ and $\Gamma_N (M_N) \approx \Gamma_N (M_{N_2})/2$  if $ \mathcal {K}^{\rm Ma}_2 \gg  \mathcal {K}^{\rm Ma}_1$, then it is not expected a significant impact if one factor dominates over the other. However, in this work we will assume that  $\mathcal{K}^{\rm Ma}_1 \approx \mathcal{K}^{\rm Ma}_2 \equiv \mathcal{K}$. In adittion, we will consider the mixing elements $|B_{\mu N_i}|^2 \approx |B_{\tau N_i}|^2 \equiv |B_{\ell N_i}|^2 = 10^{-5}$, ${\cal N}_{e i}^{\rm Ma}|B_{e N_i}|^2 \approx 0$  and ${\cal N}_{\mu i}^{\rm Ma} + {\cal N}_{\tau i}^{\rm Ma} \approx 15$; hence, $\mathcal{K}^{\rm Ma} = 15~|B_{\ell N}|^2$. As a consequence of the aforementioned, the HN total decay width are almost equals ($\Gamma_{N_1} \approx \Gamma_{N_2}$) and can be written as
\begin{equation}
\Gamma_{\rm Ma}(M_{N_i}) \equiv \Gamma_N(M_N) = 15 |B_{\ell N}|^2 \ \frac{G_F^2 M_{N}^5}{96\pi^3} \, .
\label{DNwidthappr}
\end{equation}

In Ref.~\cite{Cvetic:2015ura} it was obtained the $L$-dependent effective differential decay width considering the effect of HNOs (see Eq.~\ref{effdwfosc}) and considering the effects of a detector of lenght $L$, for fixed values\footnote{We notice that in the literature \cite{Cvetic:2014nla,Cvetic:2015naa,Cvetic:2015ura}, in the laboratory frame ($\Sigma$), usually $\gamma_N \beta_N=2$.} of HN velocity ($\equiv \beta_N$) and HN Lorentz factor ($\equiv \gamma_N$)
\begin{small}
\begin{align}
\nonumber \frac{d}{dL}\;& \Gamma(M^{\pm})  = \frac{ e^{\frac{-L \Gamma_N}{\gamma_N \ \beta_N}}}{\gamma_N \ \beta_N} \; \widetilde{\Gamma}\big( M^+ \to \ell^+_1 N \big) \ \widetilde{\Gamma}\big( N \to \ell^+_2 M^{' -}  \big)  \\
& \times \Bigg( \sum_{i=1}^2 |B_{\ell_1 N_i}|^2 |B_{\ell_2 N_i}|^2+ 2 |B_{\ell_1 N_1}| |B_{\ell_2 N_1}| |B_{\ell_1 N_2}| |B_{\ell_2 N_2}| \cos \Big(2\pi \; \frac{L}{L_{\rm osc}} \pm \theta_{LV} \Big)\Bigg) \ ,
\label{effdwfosc}
\end{align}
\end{small}
where $L_{\rm osc} = (2 \pi \beta_N \gamma_N)/\Delta M_N$ is the HN oscillation length and the angle $\theta_{LV}$  stands for the relative CP-violating phase between $N_1$ and $N_2$, that comes from the $B_{\ell N_i}$ elements\footnote{It is important to note that if $\theta_{LV}=0$, there is no difference between $\frac{d}{dL}\; \Gamma(B^{+})$ and $\frac{d}{dL}\; \Gamma(B^{-})$.} and is given by

\begin{equation}
\theta_{LV} = arg(B_{\mu N_2})+arg(B_{\tau N_2})-arg(B_{\mu N_1})-arg(B_{\tau N_1}).
\end{equation}

It is worth to mention, that, in general, $M$ is moving in the lab frame when it decays into $N$ and $\ell_1$, therefore, the product $\gamma_N \beta_N$ is not always fixed, and can be written as
\begin{equation}
\beta_N \gamma_N = \sqrt{(E_N({\hat p}'_N)/M_N)^2 - 1},
\label{bNgN}
\end{equation}
where $E_N$ is the heavy neutrino energy in the lab frame, depending on ${\hat p}'_N$ direction  in the $M$-rest frame ($\Sigma'$).
\begin{figure}[hbt]
\centering
\includegraphics[width=8cm]{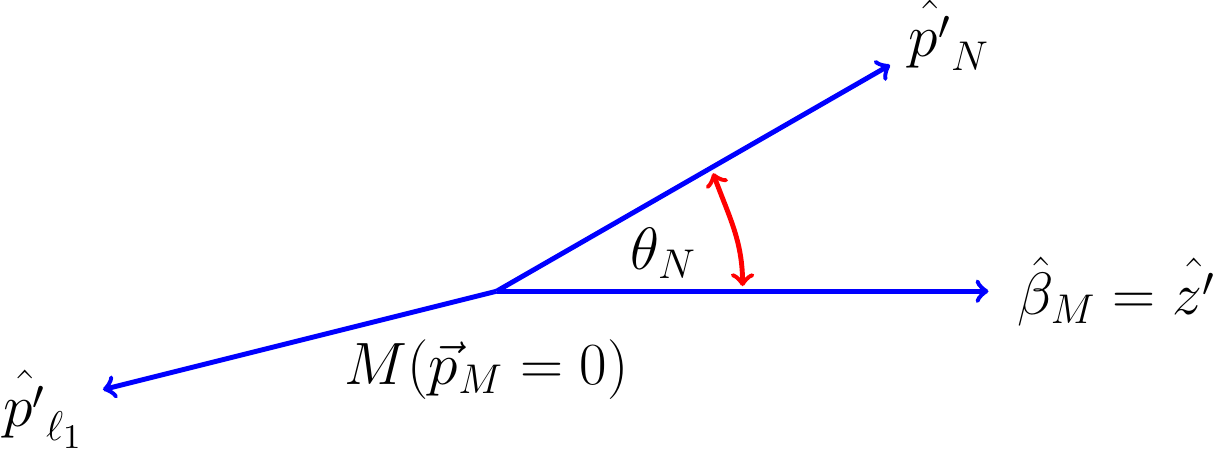}
\caption{The 3-momentum directions of leptons in the $M$-rest frame ($\Sigma'$). Here $\theta_N$ define the angle between ${\hat \beta}_M$ and ${\hat p}'_N$, where ${\hat \beta}_M=\frac{\vec{\beta}_M}{|\vec{\beta}_M|}$ is the direction of the velocity of $M$ in the lab frame, we notice that ${\hat \beta}_M$ also defines the ${\hat z}'$-axis.}
\label{FigpN}
\end{figure}

The relation among $E_N$, $\vec{p'}_N$ and the angle $\theta_N$ is given by the Lorentz energy transformation (see Fig.~\ref{FigpN})
\begin{equation}
E_N = \gamma_M (E'_N + \cos \theta_N \beta_M |{\vec p}'_N|),
\label{EN}
\end{equation}
where the corresponding factors in the $M$-rest frame ($\Sigma'$) are given by
\begin{equation}
E'_N = \frac{M_M^2 + M_N^2 - M_{\ell_1}^2}{2 M_M}, \quad
|\vec{p'}_N| = \frac{1}{2} M_M \lambda^{1/2} \left( 1, \frac{M_{\ell_1}^2}{M_M^2}, \frac{M_N^2}{M_M^2} \right),
\label{ENppNp}
\end{equation}
we remarks that $\beta_M$ is the velocity of $M$ in the lab frame, and $\lambda(x,y,z)$ is 
\begin{equation}
\lambda(x,y,z)=x^2+y^2+z^2-2xy-2xz-2yz
\end{equation}

Therefore, the Eq.~\ref{effdwfosc} must be re-written in differential form and integrated over all directions of heavy neutrino $\vec{p'}_N$ in the $M$-rest frame, in addition, we set $M \to B_c$, $\ell_1 \to \mu$, $\ell_2 \to \tau$ and $M^{'}\to \pi$
\begin{align}
\frac{d}{dL}\Gamma_{LV}^{\rm osc}(B_{c}^{\pm})&=  \int  \frac{e^{\frac{-L \Gamma_N}{\left( (E_N({\hat p}'_N)/M_N)^2 - 1 \right)^{1/2}}}}{\left( (E_N({\hat p}'_N)/M_N)^2 - 1 \right)^{1/2} } \ d \Omega_{{\hat p}'_N} \ \frac{d \widetilde{\Gamma}\big( B_c^+ \to \mu^+ N \big)}{ d \Omega_{{\hat p}'_N}} \ \widetilde{\Gamma}\big( N \to \tau^+ \pi^-  \big) \nonumber \\ 
& \times \Bigg( \sum_{i=1}^2 |B_{\mu N_i}|^2 |B_{\tau N_i}|^2+ 2 |B_{\mu N_1}| |B_{\tau N_1}| |B_{\mu N_2}| |B_{\tau N_2}| \cos \Big(2\pi \; \frac{L}{L_{\rm osc}({\hat p}'_N)} \pm \theta_{LV} \Big)\Bigg), 
\label{effdwfosc2}
\end{align}
where $L_{\rm osc}({\hat p}'_N)$ adopts the following form
\begin{equation}
L_{\rm osc}({\hat p}'_N) = \frac{2 \pi\beta_N \gamma_N}{M_N} = 
\frac{2 \pi}{M_N^2} |{\vec p}_N({\hat p}'_N)| =
\frac{2 \pi}{M_N} \left[ (E_N({\hat p}'_N)/M_N)^2 - 1 \right]^{1/2}\ ,
\label{Losc2}
\end{equation}
and 
\begin{eqnarray}
\overline{\Gamma}(N \to\pi^{\pm}  \tau^{\mp} ) & = &
\frac{1}{16 \pi} G_F^2 f_{\pi}^2 |V_{ud}|^2 \frac{1}{M_N} 
\; \lambda^{1/2}\left(1, \frac{M_{\pi}^2}{M_N^2}, \frac{M_{\tau}^2}{M_N^2}\right)\times 
\nonumber \\ 
&& \left[ \Big(M_N^2 + M_{\tau}^2\Big) \Big(M_N^2 - M_{\pi}^2+M_{\tau}^2\Big) -
4 M_N^2 M_{\tau}^2 \right]\ .
\end{eqnarray}
The term $d {\widetilde{\Gamma}} ( B^+ \to \ell_{1}^+ N )/d \Omega_{{\hat p}'_N}$ is gven by
\bes
\label{dGWmuN}
\bea
\frac{d {\widetilde{\Gamma}} ( B_{c}^+ \to \mu^+ N )}{d \Omega_{{\hat p}'_N}} &=&
\frac{1}{4 \pi} {\widetilde{\Gamma}} ( B_{c}^+ \to \mu^+ N )
\\
\nonumber
& = & \frac{1}{32 \pi^2} G_F^2 f_{B_c}^2 |V_{cb}|^2 M_{B_c}^3 \lambda^{1/2}(1,x_N,x_{\mu}) \Big( (1-x_N)x_N+x_{\mu}(1+2x_N-x_{\mu}) \Big) \\
\eea
\ees
where $x_N = M_N^2/M_{B_c}^2$ and $x_{\mu}=M_{\mu}^2/M_{B_c}^2$. The Fermi constant is $G_F=1.166 \times10^{-5}\ {\rm GeV}^{-2}$, the meson decays constants are $f_{\pi}=0.1304\ {\rm GeV}$ and $f_{B_c}=0.4\ {\rm GeV}$, the CKM elements are $|V_{ud}|=0.974$ and $|V_{cb}|=0.041$, and the masses $M_{B_c}=6.275\ {\rm GeV}$, $M_{\pi}=139.57\times10^{-3}\ {\rm GeV}$, $M_{\mu}=105.7\times10^{-3}\ {\rm GeV}$ and $M_{\tau}=1.777\ {\rm GeV}$. It is worth mentioning that in Eq.~\ref{dGWmuN} it has been performed the average over $B_c$ initial polarization and the sum over the helicities of $\mu^+$ and $N$. Therefore,  the Eq.~\ref{effdwfosc2} is then only $\theta_N$ dependent (see ~Eqs.~\ref{EN} and \ref{ENppNp}), hence, the integration $d \Omega_{{\hat p}'_N}$ reduce to $2 \pi\ d(\cos \theta_N)$.


\section{Heavy Neutrino Production Simulations}
\label{sec:simu}

\begin{figure}[hbt]
\centering
\includegraphics[scale = 0.7]{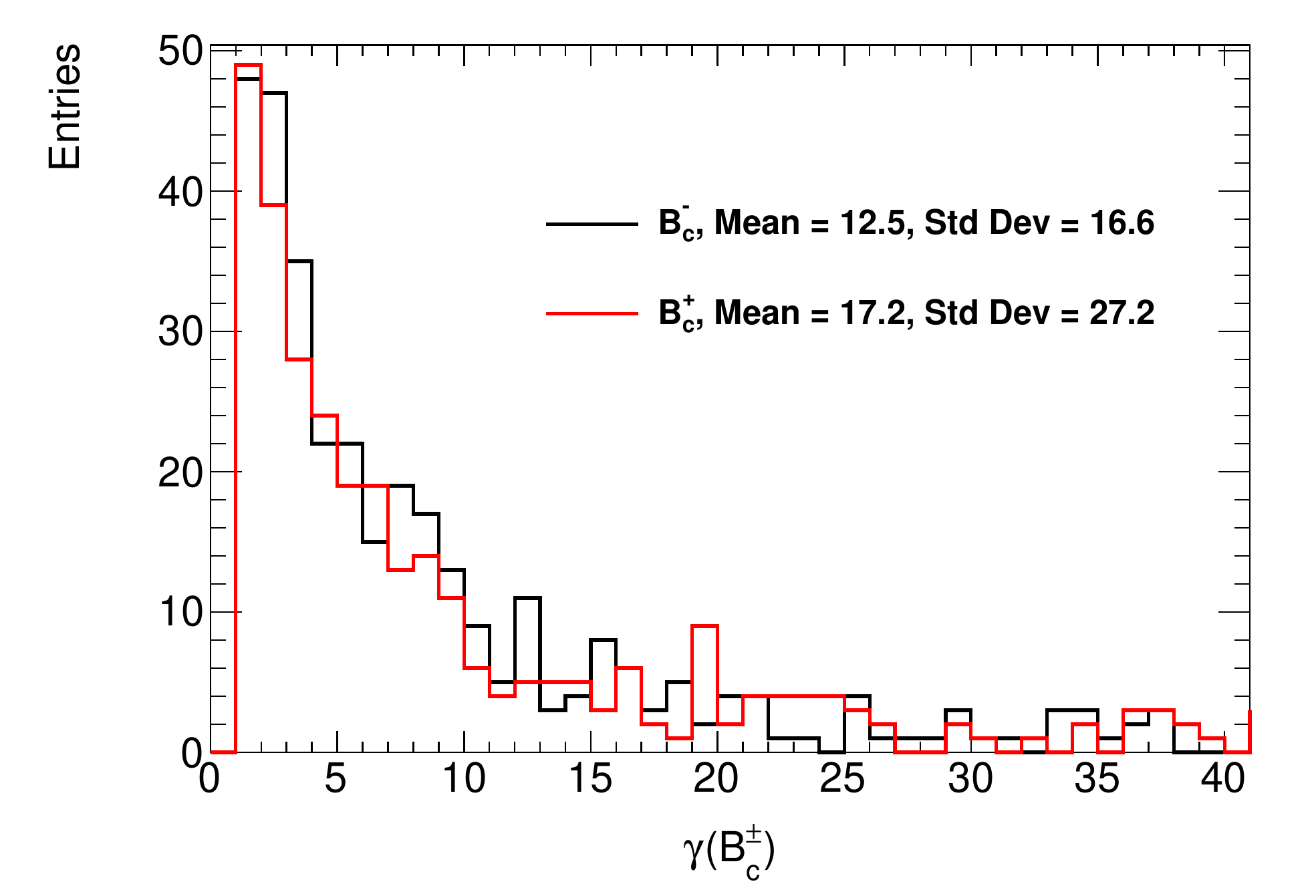}
\caption{The Lorentz $\gamma_{B_{c}^{\pm}}$ factor for $B_{c}^{\pm}$ mesons.}
\label{fig:g}
\end{figure}

For the correct evaluation of the quantities in Eq.~\ref{effdwfosc2}, and to test the feasibility to measure the phenomenon described here, we require a realistic distribution of $\gamma_M$ which can lead to a realistic distribution of $\beta_M$ through $\beta_{M}=\sqrt{1-1/\gamma_M^2}$. This realistic distribution is obtained by means of simulations of $B_c$ mesons production via charged current Drell-Yan process, using 
 \textsc{MadGraph5\_aMC@NLO}~\cite{Alwall:2014hca}, \textsc{Pythia8}~\cite{Sjostrand:2007gs} and \textsc{Delphes}~\cite{deFavereau:2013fsa} for $B_{c}^{+}$ and $B_{c}^{-}$ individually (see Fig.~\ref{fig:g}), for LHCb conditions with $\sqrt{s}=13$ TeV.

The observation of the studied phenomenon (Eq.~\ref{effdwfosc2}) depends on the number of produced $B_c$ mesons ($N_{B_c}$) at the particular experiment. The HL-LCHb is design to reach a luminosity $\mathcal{L}=2\cdot 10^{-34}\ {\rm cm^{-2} s^{-1}}$~\cite{Bediaga:2018lhg}, transforming it into one of the most promising $B$ factories. The $B$ mesons production cross-sections is $\sigma_{B}\approx86.6\ \mu b$~\cite{Aaij:2017qml}, however, $\sigma_{B_c}$ is suppressed by a factor $10^{-3}$ respect to $\sigma_{B}$ \cite{Berezhnoy:1997fp}, this suppression factor implies that for each $10^6$ $B$ mesons we have $10^3$ $B_c$ mesons. The HNs production has been calculated in detail in Refs.~\cite{Cvetic:2014nla,Cvetic:2015naa,Cvetic:2013eza}, in addittion, assuming a 50\% detector efficency the expected number of Heavy Neutrino events (with HNs masses between $3.5-5.5$ GeV and $|B_{\ell N}|^2=10^{-5}$) can reach $\approx 3000$ for 6 years of operation.

\section{Results and discussion}
\label{sec:dis}

In this article, we have studied the modulation $d\Gamma(B_c)/dL$ for the LNV $B^{\pm}_c$ meson decays assuming conditions that could be present at LHCb experiment. We focus on a scenario that contains two almost degenerate (on-shell) heavy Majorana neutrinos. This scenario has been studied in previous work Ref.~\cite{Cvetic:2015ura} in which we have explored the modulation in a more academic frame, in this paper we consider more realistic conditions that could lead to a discovery in the upcoming years. 

\begin{figure}[hbt]
\centering
\includegraphics[scale = 0.5]{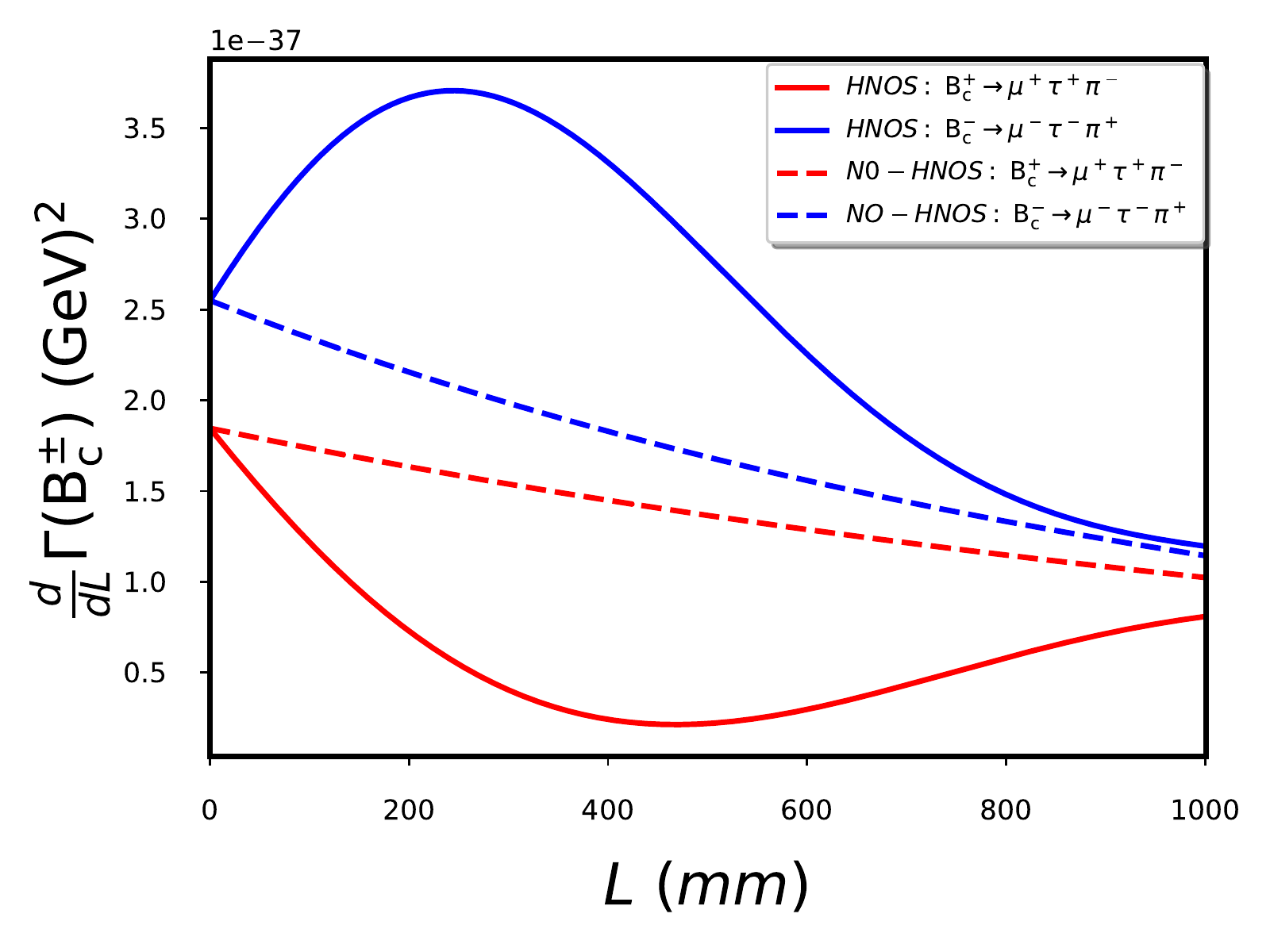}
\includegraphics[scale = 0.5]{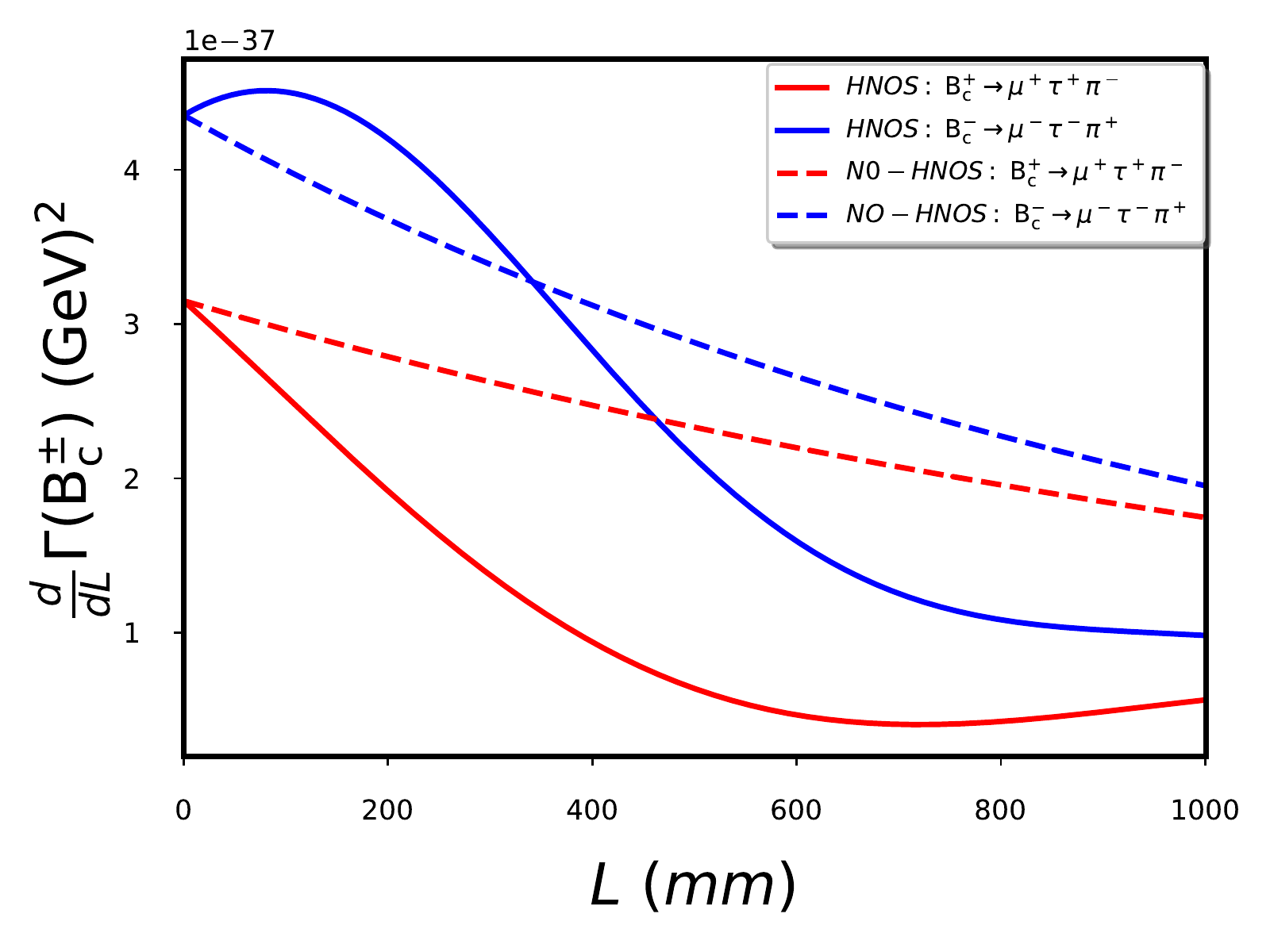}
\caption{Diferential decay width $d\Gamma(B_c)/dL$ for average value of $\gamma_{B_c^{\pm}}$. Left Panel: $M_N=3.5$ GeV, $Y=5$, $|B_{\ell N}|^2=5 \times 10^{-6}$ and $\theta_{LV}=\pi/2$. Right Panel: $M_N=3.5$ GeV, $Y=5$, $|B_{\ell N}|^2=5 \times 10^{-6}$ and $\theta_{LV}=\pi/4$. Solid lines stand for processes including the HNOS effects, while the dashed ones stand for process with NO-HNOS effects (only amplitude interference effects).}
\label{fig:5}
\end{figure}

The Fig.~\ref{fig:5} shows the Differential Decay Width $d\Gamma(B^{\pm}_c)/dL$ for fixed values of $\gamma_N$ and $\beta_N$, which are determined from the average values of $\gamma_{B^{\mp}_c}$ presented in Fig.~\ref{fig:g}, for two values of $\theta_{LV}$.  The solid lines stand for the processes which include the effects of HNOS, while the dashed lines do not. It could be seen that the effects of HNOS over   $d\Gamma(B^{\pm}_c)/dL$  could enhance or decrease it near a factor of two in comparison with the case with NO-HNOS,  for some regions of $L$. In addition, for $d\Gamma(B^{\pm}_c)/dL$ with NO-HNOS effects there is no modulation and only it is present the damped effect produced due to the probability that the HN decay. 

We noticed, that the difference between the process for $B_c^+$ and $B_c^-$ is maximized when the CP violation angle is $\theta_{LV} = \pi/2$ (as expected from Eq.~\Ref{effdwfosc2}). We can also observed that as the distance $L$ grows, both curves tend to converge, this is because as the HN propagates, the cumulative probability that the HN has decayed is greater, this effect is characterized by the exponential factor present in $d\Gamma(B_c)/dL$ (Eq.~\Ref{effdwfosc2}), which specifically accounts for the probability that the HN decays within the detector of length L.

\begin{figure}[hbt]
\centering
\includegraphics[scale = 0.5]{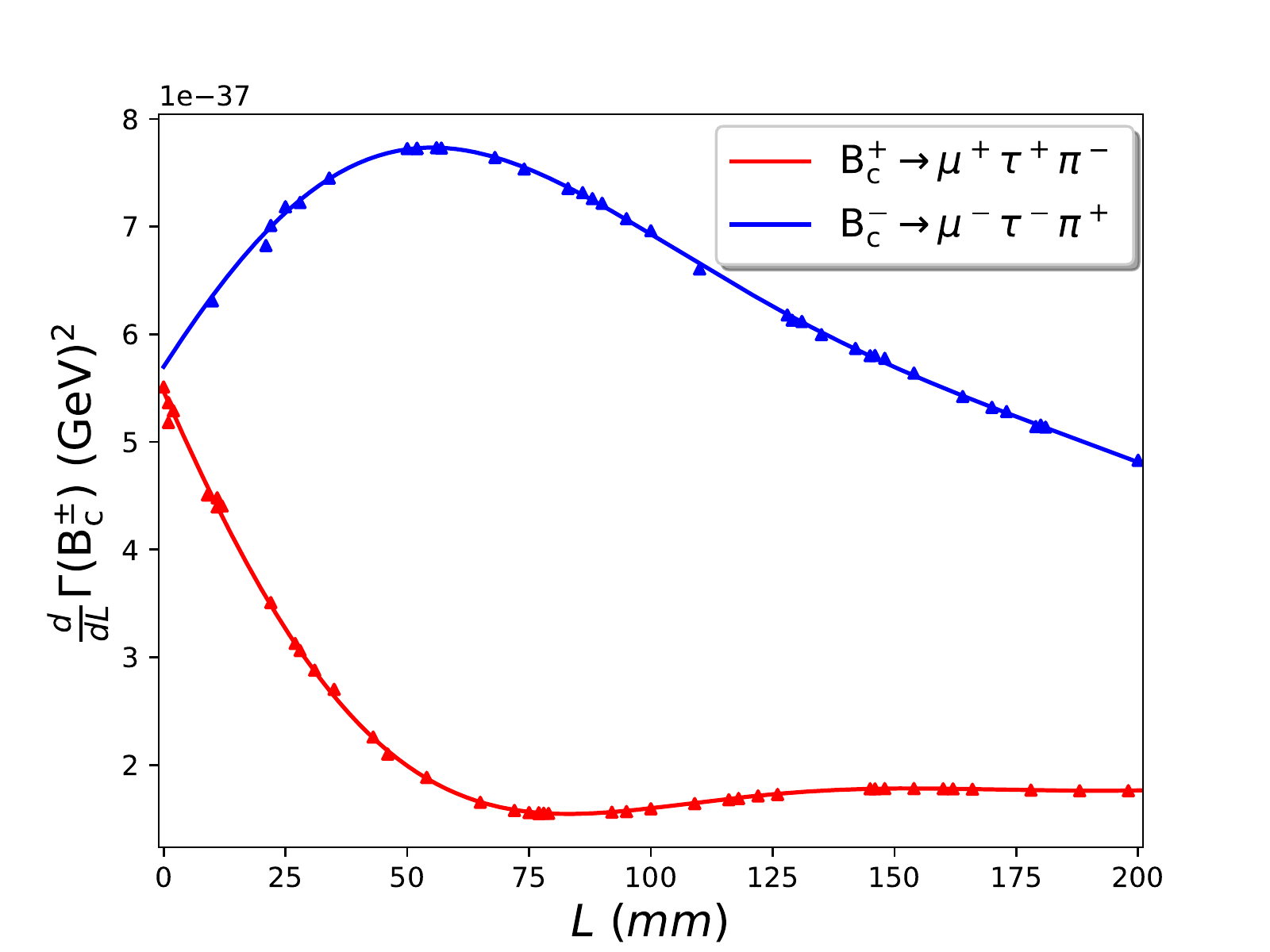}
\includegraphics[scale = 0.5]{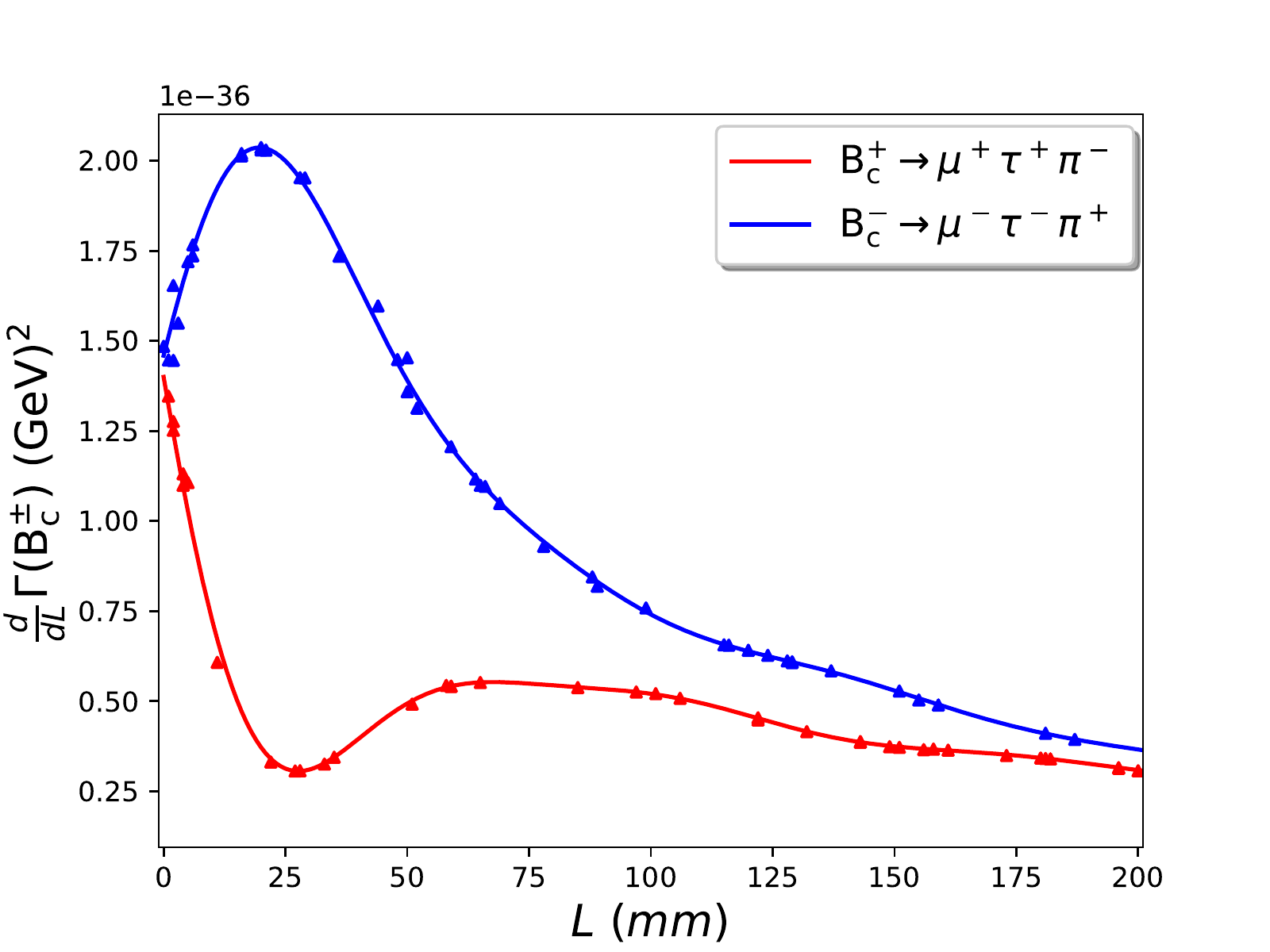}
\caption{Diferential decay width $d\Gamma(B_c)/dL$ for $\gamma_{B_c^{\pm}}$ distributed according Fig.~\ref{fig:g}. Left Panel: $M_N=3.5$ GeV, $Y=5$, $|B_{\ell N}|^2=5 \times 10^{-6}$ and $\theta_{LV}=\pi/2$. Right Panel: $M_N=4.5$ GeV, $Y=5$, $|B_{\ell N}|^2=10^{-5}$ and $\theta_{LV}=\pi/2$. The solid lines are generated assuming pefect detector resolution, while triangles stand for 50 samples of $d\Gamma(B_c)/dL$ convolved with the detector resolution Reso(L) = 1.0 mm.}
\label{fig:6}
\end{figure}

\begin{figure}[hbt]
\centering
\includegraphics[scale = 0.5]{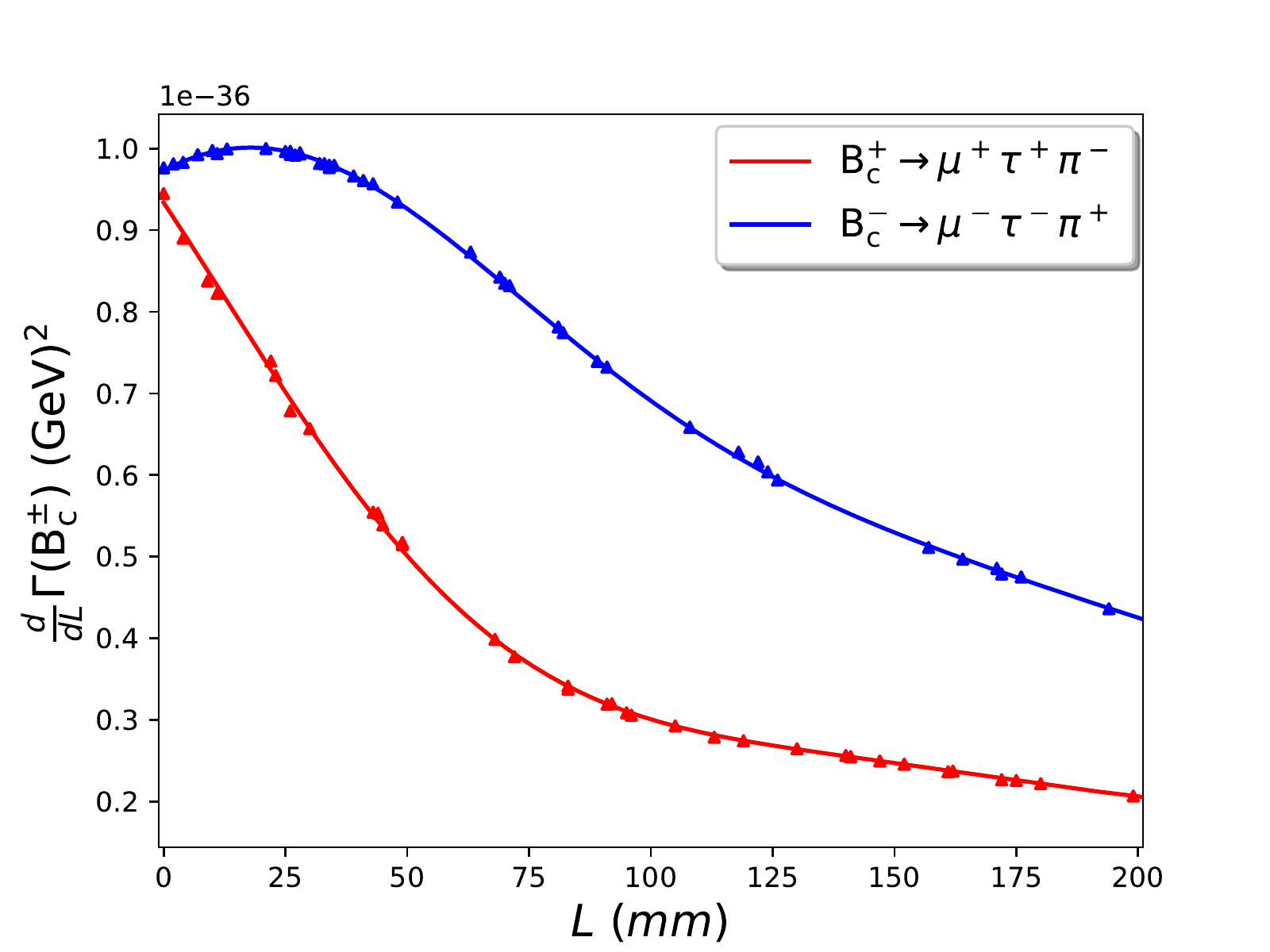}
\includegraphics[scale = 0.5]{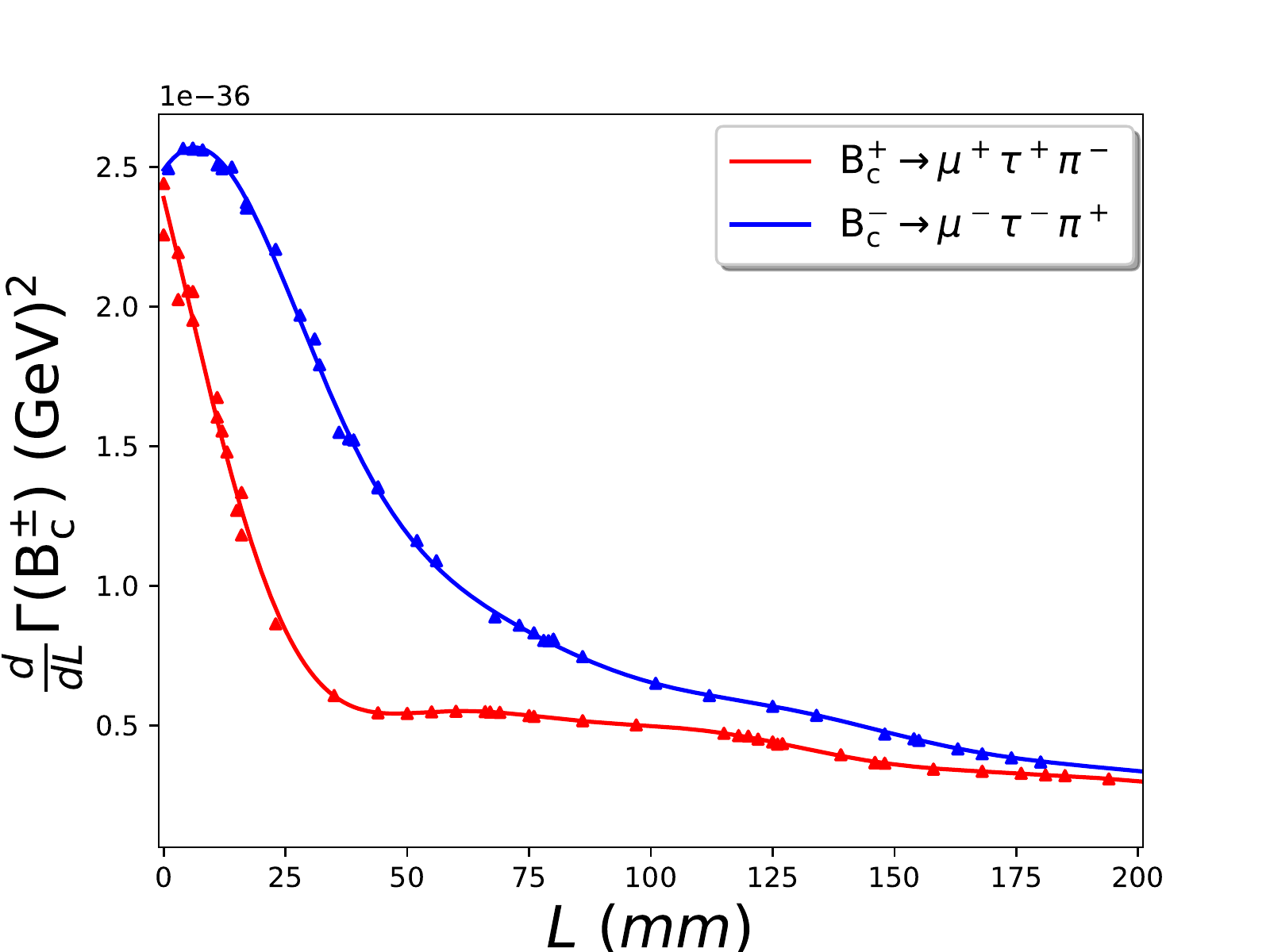}
\caption{Diferential decay width $d\Gamma(B_c)/dL$ for $\gamma_{B_c^{\pm}}$ distributed according Fig.~\ref{fig:g}. Left Panel: $M_N=3.5$ GeV, $Y=5$, $|B_{\ell N}|^2=5 \times 10^{-6}$ and $\theta_{LV}=\pi/4$. Right Panel: $M_N=4.5$ GeV, $Y=5$, $|B_{\ell N}|^2=5 \times 10^{-6}$ and $\theta_{LV}=\pi/4$. The solid lines are generated assuming pefect detector resolution, while triangles stand for 50 samples of $d\Gamma(B_c)/dL$ convolved with the detector resolution Reso(L) = 1.0 mm.}
\label{fig:7}
\end{figure}

Figs.~\Ref{fig:6} and \ref{fig:7} shows the Differential decay width $ d\Gamma(B_c)/dL$ for non-fixed  values (non-average values) of $\gamma_N$ and $\beta_N$, which are determined from the simulated distributions of $\gamma_{B^{\pm}_c}$ presented in figure Fig.~\ref{fig:g}. The Figs.~\Ref{fig:6} and \ref{fig:7} were performed for $\theta_{LV}=\pi/2$  and $\theta_{LV}=\pi/4$, respectively, and two values of $M_{N}$. The effects of non-fixed $\gamma_N$ and $\beta_N$ were calculated taking into account the relative probability of each bin of them, that is, considering the specific contribution of each bin in the final values of $d\Gamma(B^{\pm}_c)/dL$. The Figs.~\Ref{fig:6} and \ref{fig:7} also shows the results including the detector position resolution, Reso(L)=1 mm, for a sample of 50 events simulated from the $d\Gamma(B^{\pm}_c)/dL$ distributions, this effect is shown by triangles that do not fit perfectly on the continuous curve, which corresponds to the detector with perfect resolution, Reso(L)=0.0 mm. In addition, the effects of non-fixed $\gamma_N$ and $\beta_N$ are manifest by mean of a smoothing of the modulation, it can be easily seen by an eyeball comparison between Fig.~\ref{fig:5} with Figs.~\ref{fig:6} and \ref{fig:7} (left-panels).

By comparing left and right panels from Figs.~\Ref{fig:6} and \Ref{fig:7}, we can see that the maximum of the difference between the curves runs to the left, e.i in right panels the CP-violation is maximized when $L \approx 23$ mm while in left panels it maximize at $L \approx 66$ mm. This is mainly due to the fact that the larger $M_N$ implies a shorter lifetime and consequently the HN decay in shorter distances.

\begin{figure}[hbt]
\centering
\includegraphics[scale = 0.5]{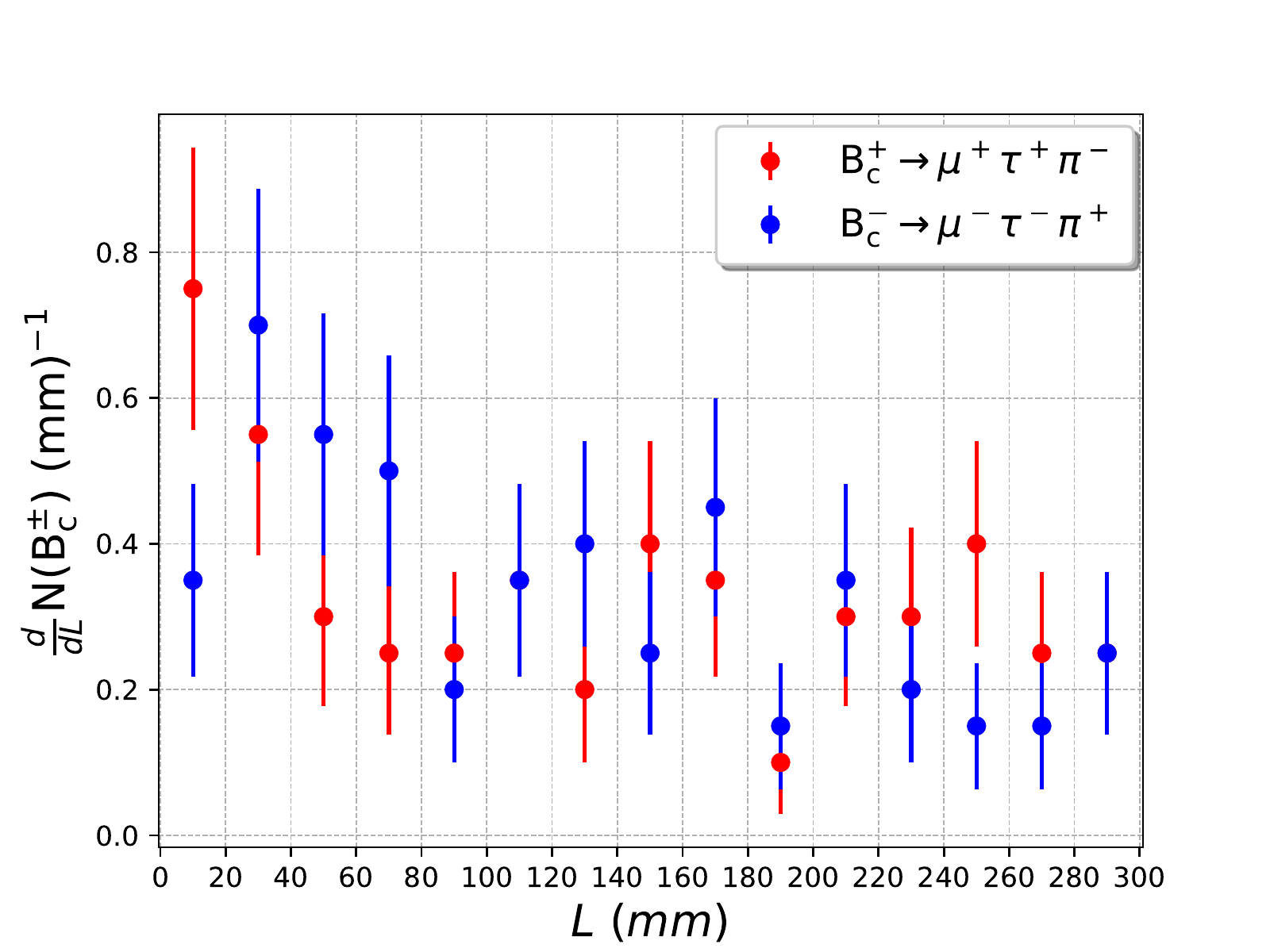}
\includegraphics[scale = 0.5]{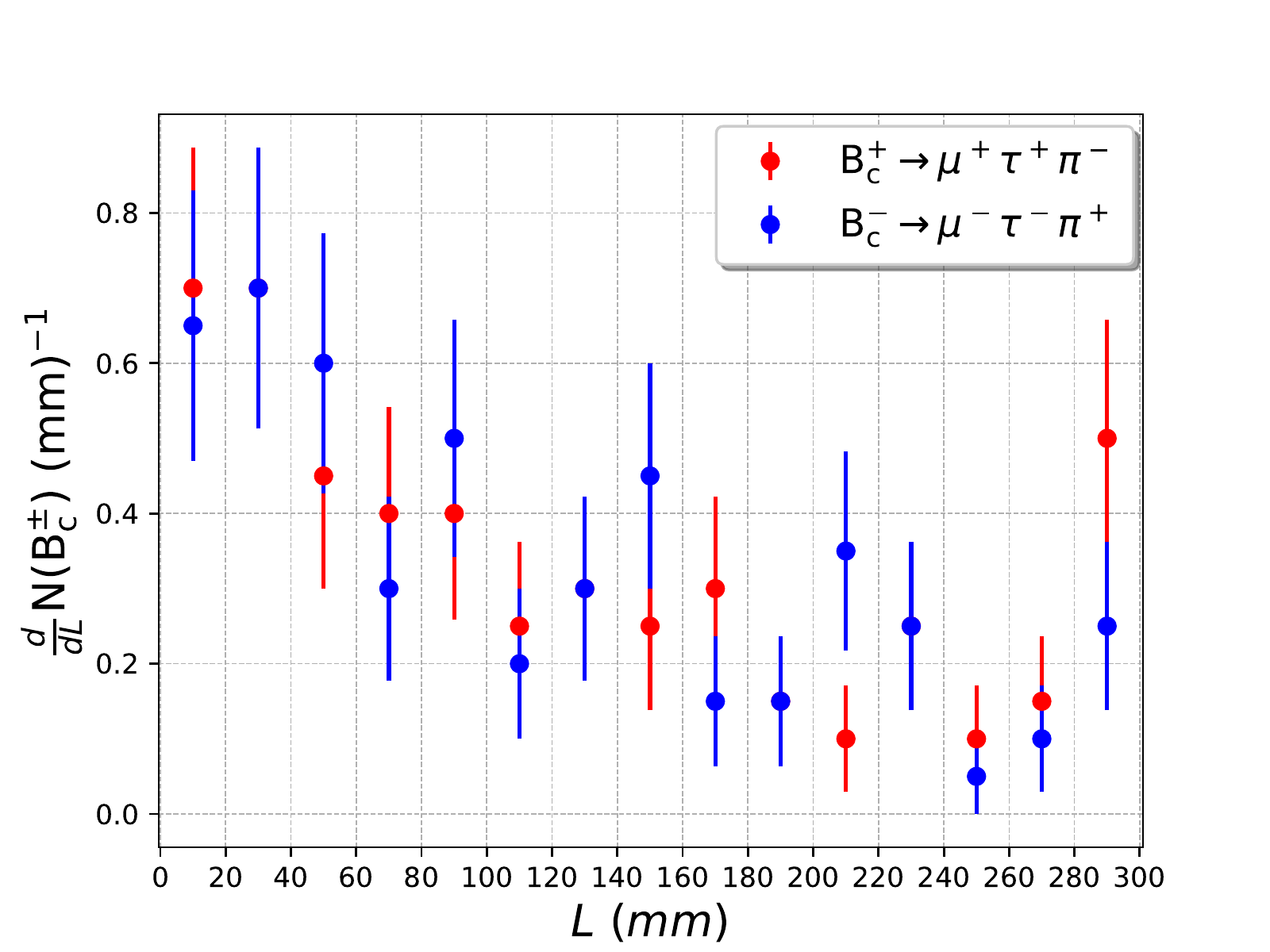}
\caption{Diferential number of events $dN/dL$ for 100 samples. Left Panel: $M_N=3.5$ GeV, $Y=5$, $|B_{\ell N}|^2=5 \times 10^{-6}$ and $\theta_{LV}=\pi/2$. Right Panel: $M_N=3.5$ GeV, $Y=5$, $|B_{\ell N}|^2=5 \times 10^{-6}$ and $\theta_{LV}=\pi/4$. Here the bin width is ${\rm \Delta L=20\ mm}$, in addition, it was considered that  $\gamma_{B_c^{\pm}}$ are distributed according Fig.~\ref{fig:g}.}
\label{fig:8}
\end{figure}

\begin{figure}[hbt]
\centering
\includegraphics[scale = 0.5]{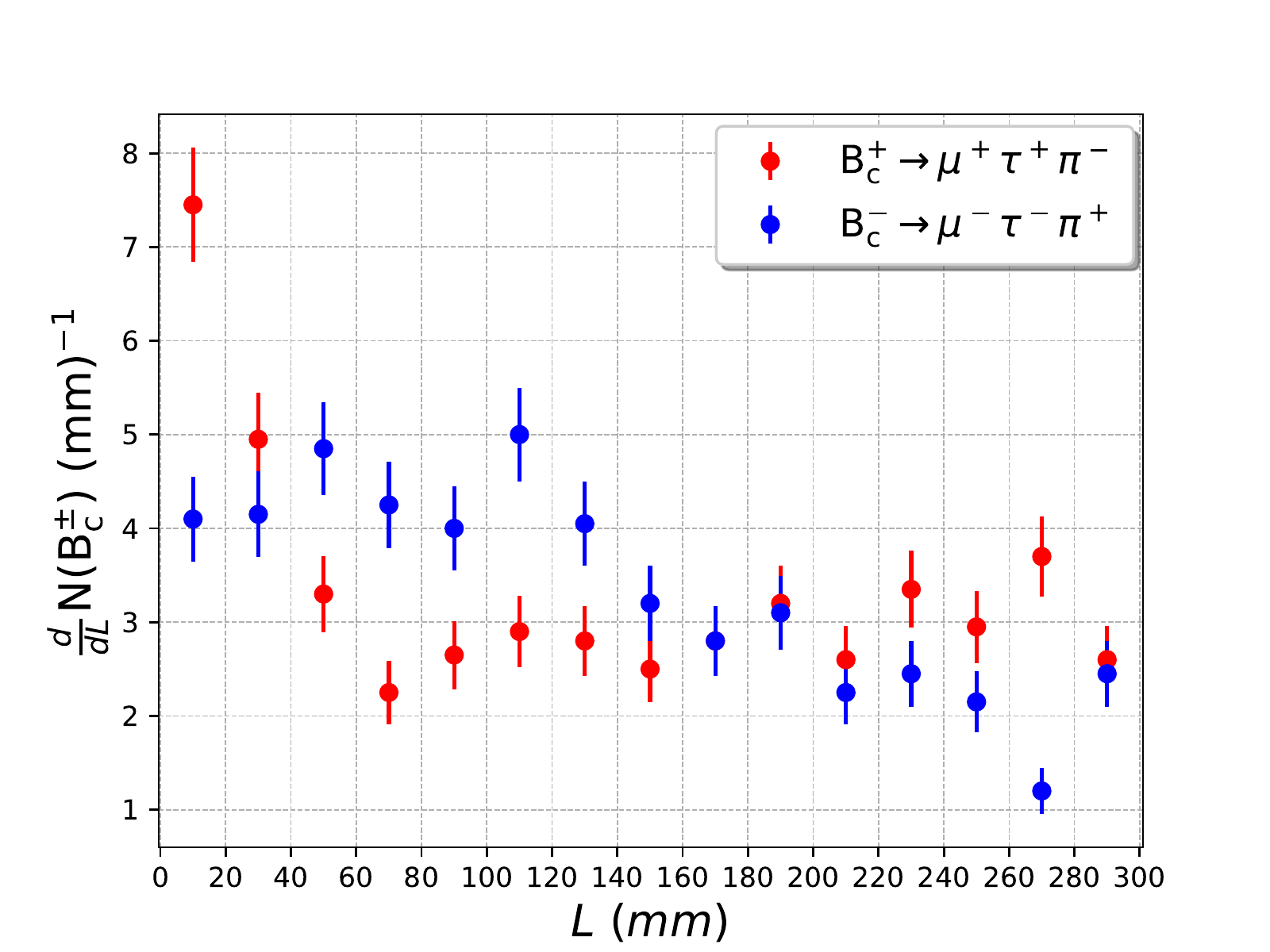}
\includegraphics[scale = 0.5]{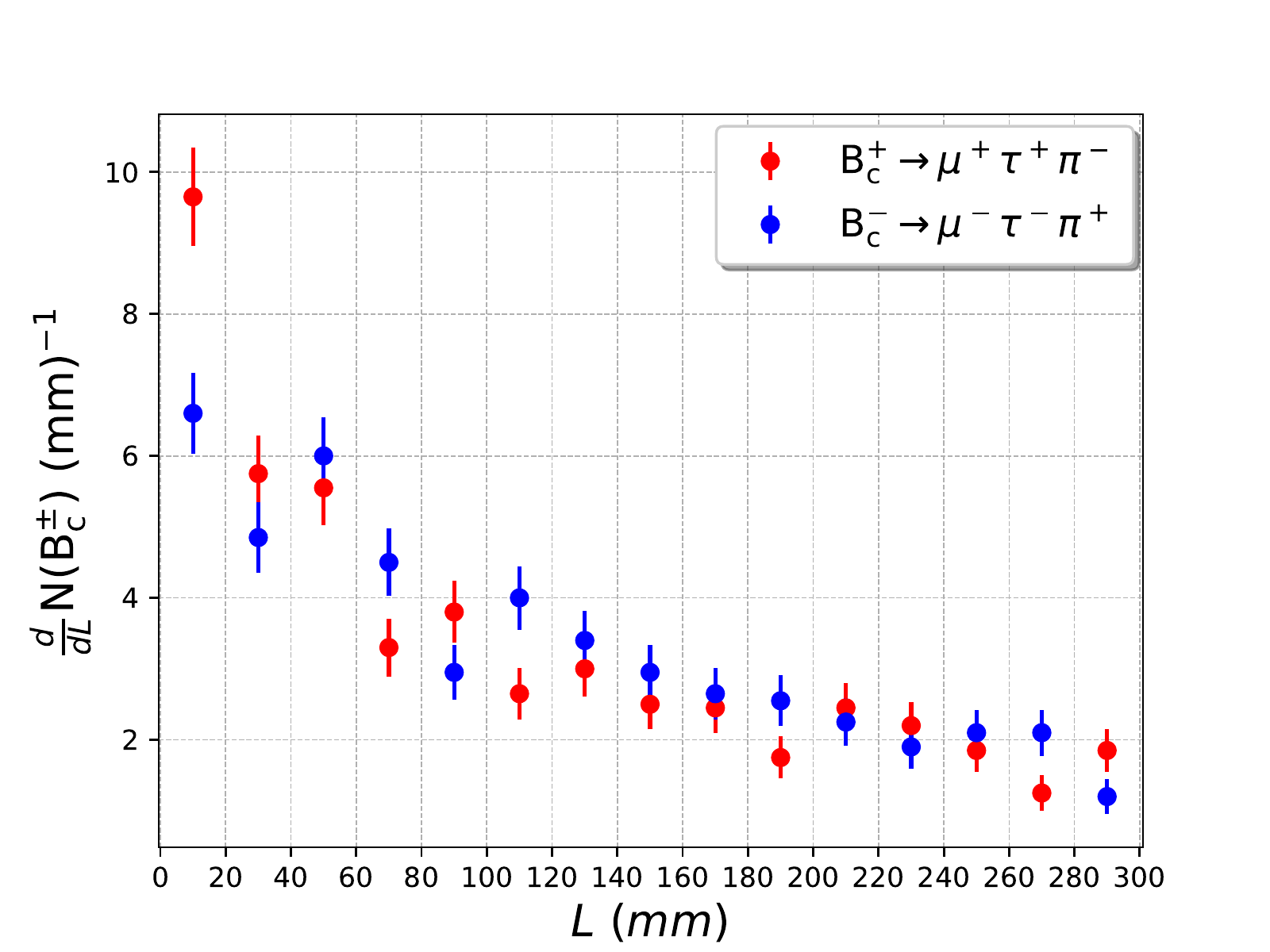}
\caption{Diferential number of events $dN/dL$ for 1000 samples. Left Panel: $M_N=3.5$ GeV, $Y=5$, $|B_{\ell N}|^2=5 \times 10^{-6}$ and $\theta_{LV}=\pi/2$. Right Panel: $M_N=3.5$ GeV, $Y=5$, $|B_{\ell N}|^2=5 \times 10^{-6}$ and $\theta_{LV}=\pi/4$. Here the bin width is ${\rm \Delta L=20\ mm}$, in addition, it was considered that  $\gamma_{B_c^{\pm}}$ are distributed according Fig.~\ref{fig:g}.}
\label{fig:9}
\end{figure}

The Figs.~\ref{fig:8} and \ref{fig:9} show the simulated, $dN(B^{\pm}_c)/dL$, distribution for $ M_N=3.5$, $Y=5$, $|B_ {\ell N}|^2=5\times10^{-6}$ GeV and two values of $\theta_{LV}$, from a sample of 100 and 1000 events, respectively. We remarks that we had simulated the same number of events for processes that involes $B_c^+$ and its CP conjugate ($B_c^-$), despide that from Eq.~\ref{effdwfosc} we knows that cross sections of $B_c^+$ and $B_c^-$ are different if $\theta_{LV} \neq 0$. Both cases include their respective statistical error and consider $\gamma_N$ and $\beta_N$ distributed according to the result presented in Fig.~\Ref{fig:g}. From Fig.~\Ref{fig:8} we can see that there is only a modest difference between $B^{\pm}_{c}$ distributions, e.g in  $L= 0 - 60$ mm, based on what we think it won't be possible to distinguish the oscillation for neither, $\theta_{LV}=\pi/2$ or $\theta_{LV}=\pi/4$, with more than $5\sigma$'s from the "non-oscillation" scenario with only 100 signal events.
A more positive scenario is show in Fig.~\Ref{fig:9}, assuming 1000 signal events, where we can clearly observed the oscillation with good precision in the whole range for $\theta_{LV}=\pi/2$, and in $L=0 - 120$ mm, $L=180 - 200$ mm and $L=260 - 300$ mm for $\theta_{LV}=\pi/4$.

\begin{figure}[hbt]
\centering
\includegraphics[scale = 0.5]{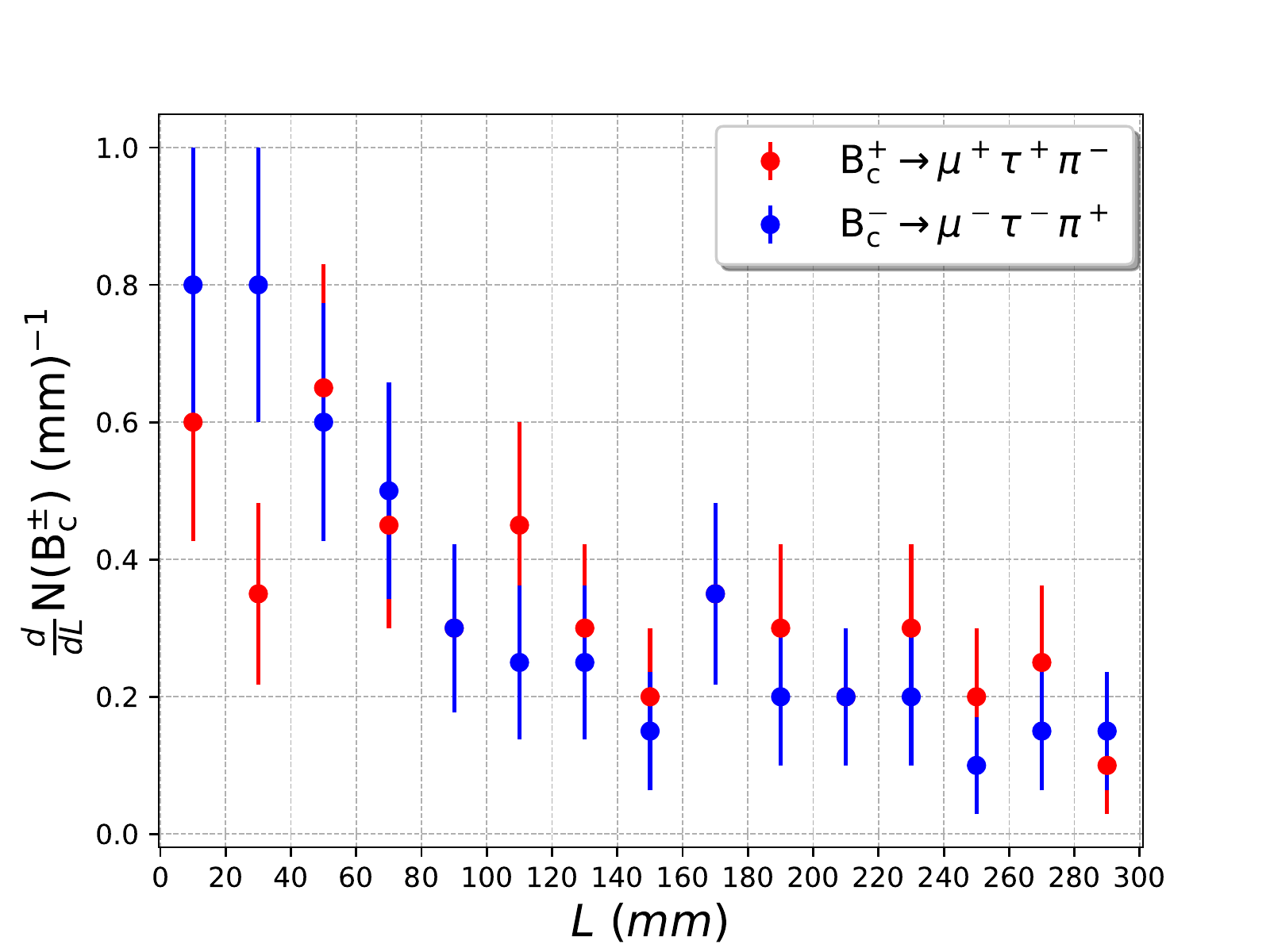}
\includegraphics[scale = 0.5]{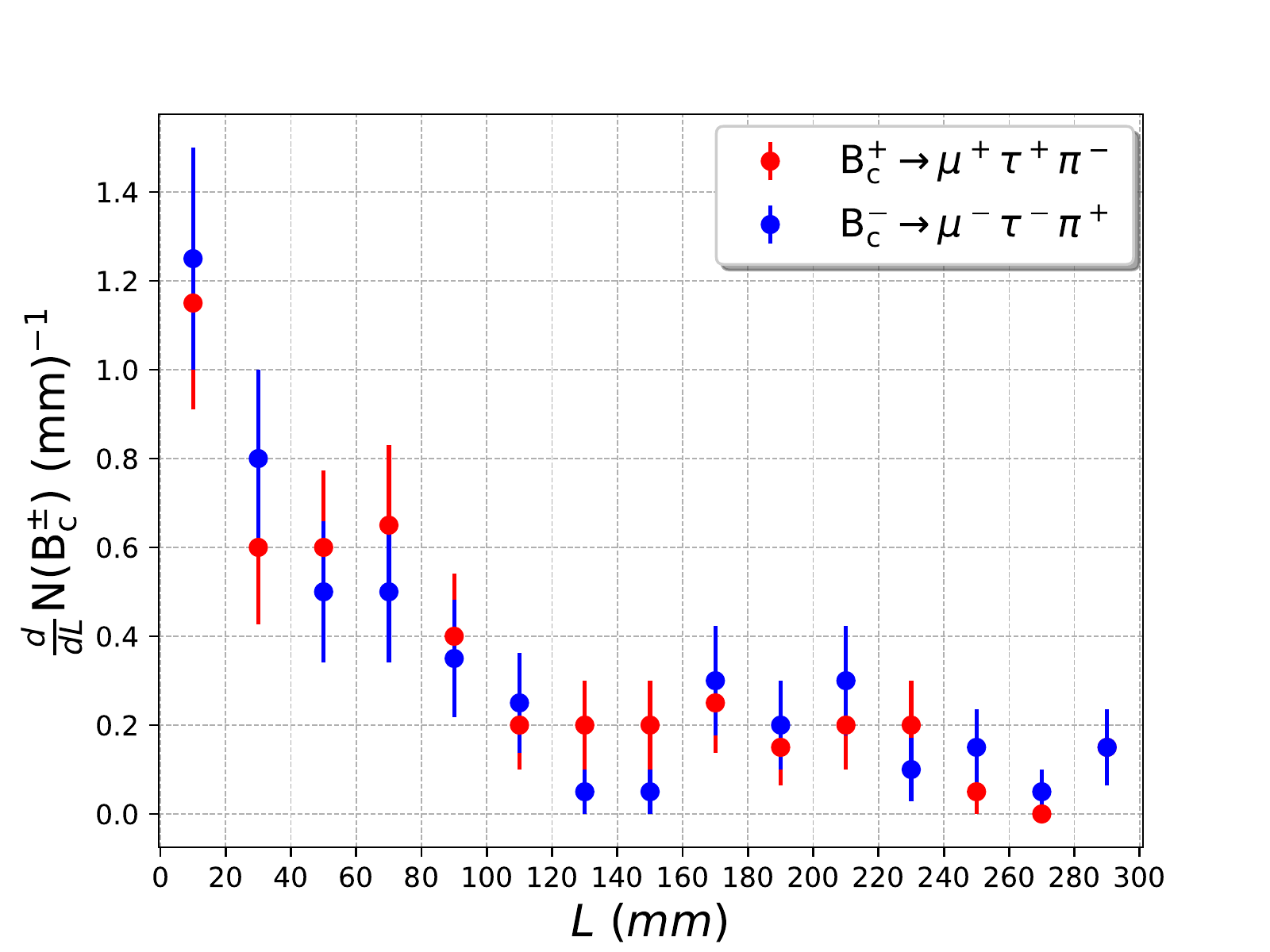}
\caption{Diferential number of events $dN/dL$ for 100 samples. Left Panel: $M_N=4.5$ GeV, $Y=5$, $|B_{\ell N}|^2=5 \times 10^{-6}$ and $\theta_{LV}=\pi/2$. Right Panel: $M_N=4.5$ GeV, $Y=5$, $|B_{\ell N}|^2=5 \times 10^{-6}$ and $\theta_{LV}=\pi/4$. Here the bin width is ${\rm \Delta L=20\ mm}$, in addition, it was considered that  $\gamma_{B_c^{\pm}}$ are distributed according Fig.~\ref{fig:g}.}
\label{fig:10}
\end{figure}

\begin{figure}[hbt]
\centering
\includegraphics[scale = 0.5]{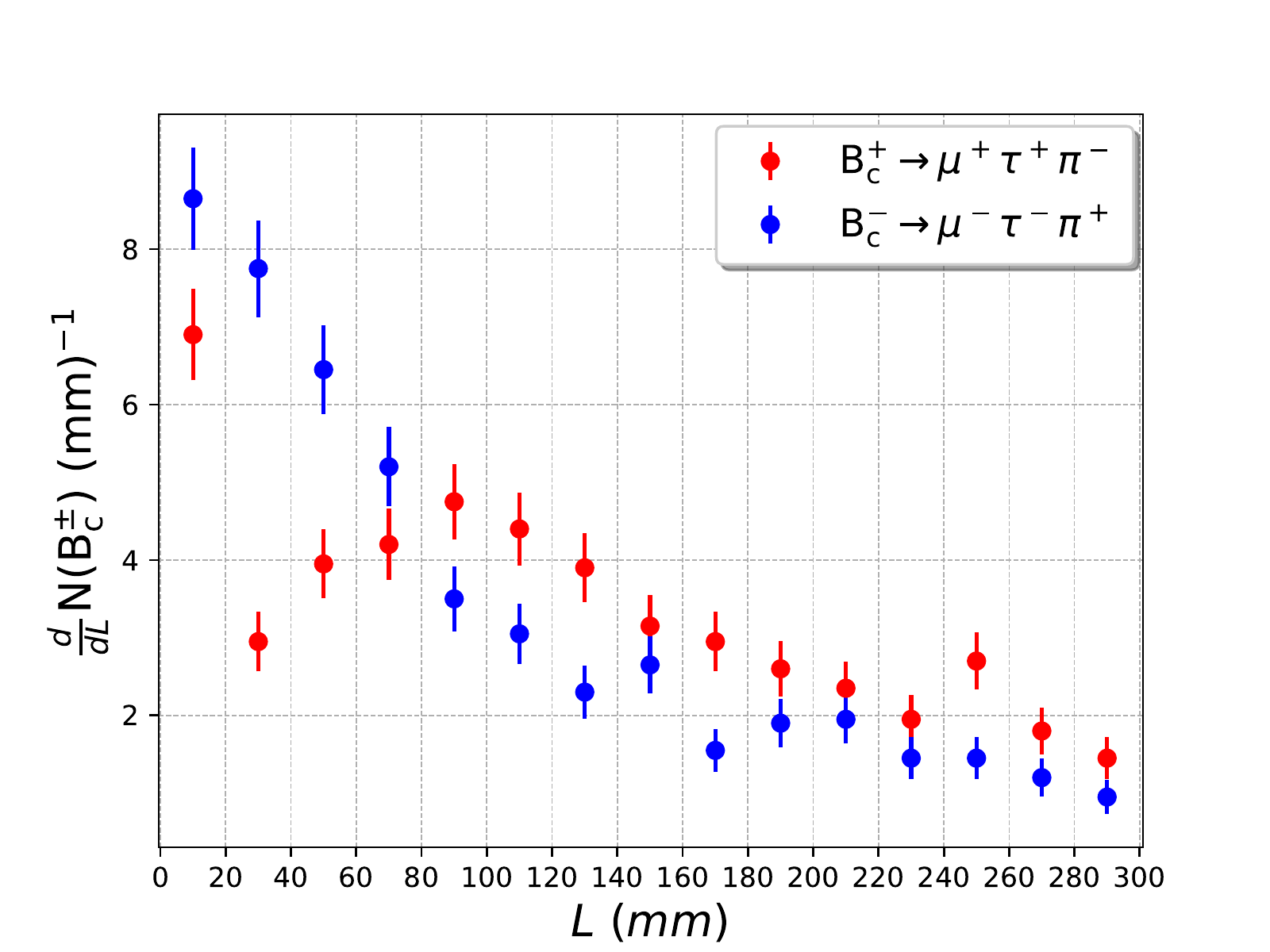}
\includegraphics[scale = 0.5]{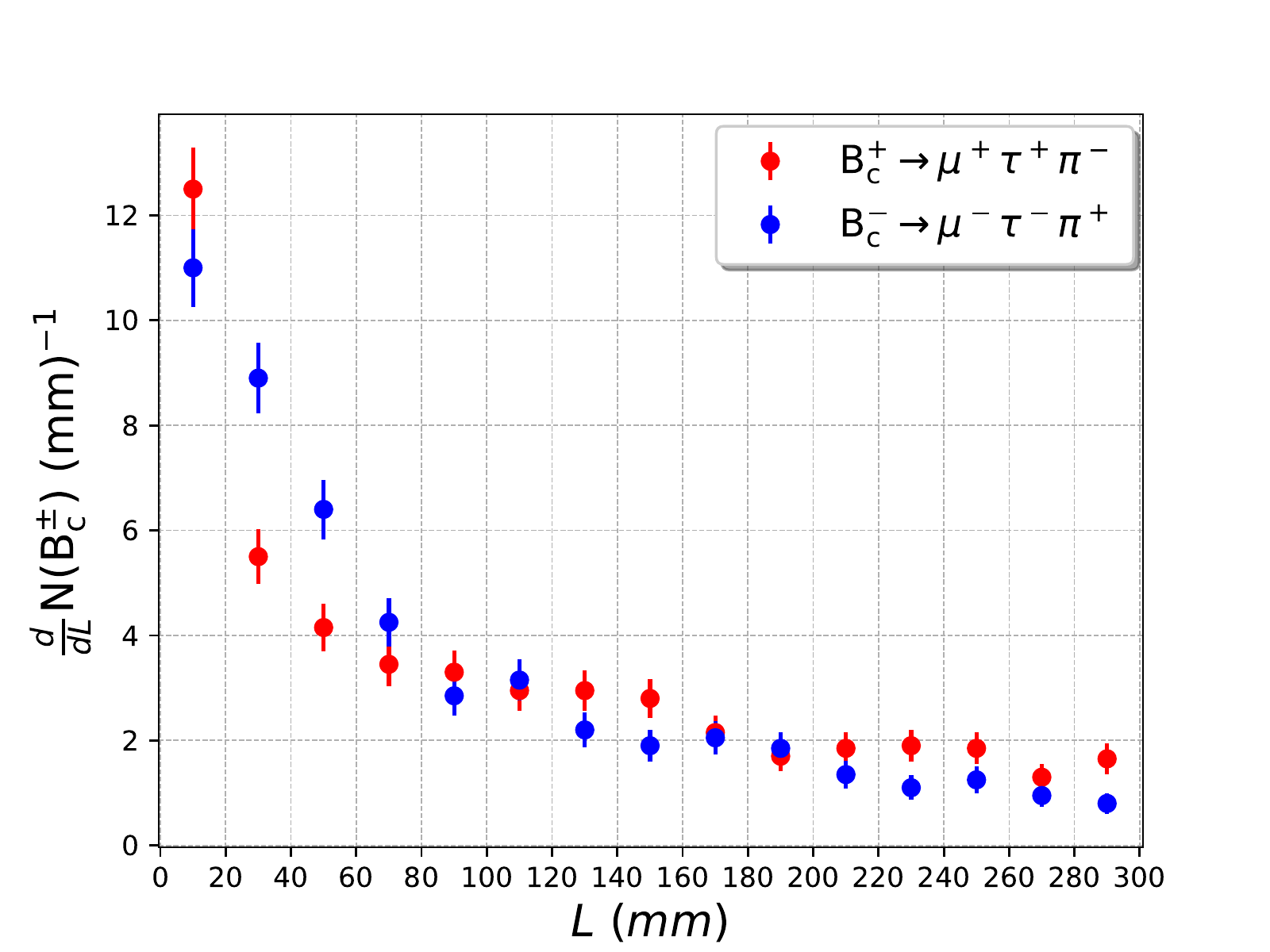}
\caption{Diferential number of events $dN/dL$ for 1000 samples. Left Panel: $M_N=4.5$ GeV, $Y=5$, $|B_{\ell N}|^2=5 \times 10^{-6}$ and $\theta_{LV}=\pi/2$. Right Panel: $M_N=4.5$ GeV, $Y=5$, $|B_{\ell N}|^2=5 \times 10^{-6}$ and $\theta_{LV}=\pi/4$. Here the bin width is ${\rm \Delta L=20\ mm}$, in addition, it was considered that  $\gamma_{B_c^{\pm}}$ are distributed according Fig.~\ref{fig:g}.}
\label{fig:11}
\end{figure}

The Figs.~\ref{fig:10} and \ref{fig:11} show the, $dN(B^{\pm}_c)/dL$, distribution for $ M_N=4.5$, $Y=5$, $|B_{\ell N}|^2=5 \times 10^{-6}$ GeV and two values of $\theta_{LV}$, from a sample of 100 and 1000 events, respectively. Both cases include their respective statistical error and consider $\gamma_N$ and $\beta_N$ distributed according to the result presented in Fig.~\Ref{fig:g}. 
For this larger mass scenario we observed that the feasibility of discover HN oscillation is possible in the whole range of L for $\theta_{LV}=\pi/2$ and for $L=20 - 60$ mm, $L=140 - 160$ mm, $L=220 - 260$ mm and $L=280 - 300$ mm  for $\theta_{LV}=\pi/4$, with a similar conclusion about statistics, addressing the $5\sigma$'s only in the 1000 signal events case.

\section{Summary}
\label{summ}

In this work we have studied the decay of HN's and their modulation in rare $B_c$ meson decays at the HL-LHCb conditions. Here we have found that the modulation produce by the HNO's could be observed if 1000 HN events are detected, this number is consistent with the expected number of HN decays at HL-LHCb.


\section{Acknowledgments}
The work of J.Z-S. and M.V-B. was funded by ANID - Millennium Program - ICN2019\_044. The work of S.T. acknowledges support from the Department of Energy, Office of Science, Nuclear Physics, under Grant DE-FG0292ER40962.

\bibliographystyle{apsrev4-1}
\bibliography{biblio.bib}
\end{document}